\documentclass[aps,prl,twocolumn]{revtex4}
\usepackage{float}
\usepackage{hyperref}
\usepackage{graphicx}
\usepackage[utf8]{inputenc}
\DeclareGraphicsExtensions{.png,.jpg,.eps,.pdf}
\usepackage{amsmath}
\usepackage{comment}
\usepackage{bm}
\usepackage{braket}
\usepackage{dsfont}
\usepackage{amsfonts}
\usepackage{xcolor}
\usepackage{bbold}
\usepackage{qcircuit}
\usepackage{physics}
\usepackage[normalem]{ulem}

\newcommand{\redmark}[1]{\color{red}\textbf{#1}\color{black}}

\begin{document}


\title{Quantum circuits to measure scalar spin chirality}

\author{L. I. Reascos$^{1}$, Bruno Murta$^{1,2}$, E. F. Galv\~ao$^{2,3}$, J. Fern\'{a}ndez-Rossier$^2$\footnote{On permanent leave from Departamento de F\'{i}sica Aplicada, Universidad de Alicante, 03690 San Vicente del Raspeig, Spain}$^,$\footnote{joaquin.fernandez-rossier@inl.int} }

\affiliation{$^1$Centro de F\'{i}sica das Universidades do Minho e do Porto, Universidade do Minho, Campus de Gualtar, 4710-057 Braga, Portugal }

\affiliation{
$^2$International Iberian Nanotechnology Laboratory (INL), Av. Mestre Jos\'{e} Veiga, 4715-330 Braga, Portugal }

\affiliation{$^3$Instituto de F\'isica, Universidade Federal Fluminense, Av. Gal. Milton Tavares de Souza s/n, Niter\'oi, RJ, 24210-340, Brazil}

\date{\today}


\begin{abstract}


The scalar spin chirality is a three-body physical observable that plays an outstanding role both in classical magnetism, characterizing non-coplanar spin textures, and in quantum magnetism, as an order parameter for chiral spin liquids. In quantum information, the scalar spin chirality is a witness of genuine tripartite entanglement. Here we propose an indirect measurement scheme, based on the Hadamard test, to estimate the scalar spin chirality for general quantum states. We apply
our method to study chirality in two types of quantum states: generic one-magnon states of a ferromagnet, and the ground state of a model with competing symmetric and antisymmetric exchange. We show a single-shot determination of the scalar chirality is possible for chirality eigenstates, via quantum phase estimation with a single auxiliary qutrit. Our approach provides a unified theory of chirality in classical and quantum magnetism.
\end{abstract}

\maketitle

Digital quantum simulation holds the promise of addressing many-body problems that are out of reach of conventional numerical methods~\cite{Cao19, McArdle20}. Thanks to exponential algorithmic speed-ups~\cite{Lloyd96, Childs18} 
and the natural encoding of entanglement, 
gate-based quantum computers are expected to allow for an efficient representation of correlated quantum states of matter. Once such a state is prepared on quantum hardware, quantum routines have to be devised to extract useful information about its physical properties. The development of these methods should minimize the number of state preparation instances and measurements as the number of relevant observables often scales with the system size and the emergent nature of quantum many-body phenomena~\cite{Anderson72} demands the study of sufficiently large systems. 

Different approaches can be adopted to determine the expectation values of physical quantities from the output of a quantum circuit. One approach is the reconstruction of its density matrix, either in full through quantum state tomography~\cite{NielsenChuang10} or approximately 
via classical shadows~\cite{Huang20}. Such an exhaustive characterization of the output state is only justified if multiple expectation values have to be computed. When one targets a single observable, the explicit estimation of its expectation value is the natural approach, namely by expanding such Hermitian operator in a complete basis of unitary operators (e.g., the Pauli basis) and measuring the expectation value of each unitary term directly~\cite{Peruzzo14}. An alternative approach to this direct measurement is encoding the Hermitian operator in a unitary operation, applying it subject to the control of ancillary qubits to the reference state, and then measuring the ancillary qubits to retrieve the expectation value, possibly requiring post-processing. The canonical example of such an indirect measurement procedure is the Hadamard test~\cite{Cleve98}, which involves a single ancillary qubit. Quantum phase estimation (QPE)~\cite{Kitaev95, NielsenChuang10} generalizes it to multiple ancillary qubits 
and can prepare exact eigenstates of a Hermitian operator~\cite{NielsenChuang10}.

The route taken here is to devise indirect measurement schemes to determine the expectation value of the scalar spin chirality for trios of spins-$\frac{1}{2}$ in quantum spin states of arbitrary sizes encoded on a quantum computer, thus avoiding the direct measurement protocol, which takes more measurements to achieve a given precision. The scalar spin chirality is the three-spin-$\frac{1}{2}$ observable~\cite{wen1989}
\begin{equation}
    \hat{\chi}_{i,j,k} = \frac{4}{\sqrt 3} \; \vec{S}_i\cdot\left(\vec{S}_j\times\vec{S}_k\right).
    \label{eq:chirality}
\end{equation}
$\vec{S}_i \equiv \frac{1}{2}(\sigma^{x}_{i}, \sigma^{y}_{i}, \sigma^{z}_{i})$ are the spin-$\frac{1}{2}$ operators at site $i$, where ($\sigma^x$, $\sigma^y$, $\sigma^z$) are the Pauli matrices. 
The prefactor ensures that its eigenvalues are $\lambda = 0, \pm 1$. 
The eigenvectors of $\hat{\chi}$ include $\ket{\uparrow \uparrow \uparrow}$, with $\lambda = 0$, and 
\begin{eqnarray}
    \frac{1}{\sqrt{3}} \Big( \ket{\uparrow \downarrow \downarrow} + \omega_{\lambda} \ket{\downarrow \uparrow \downarrow} + \omega_{\lambda}^2  \ket{\downarrow \downarrow \uparrow} \Big),
\label{eq:pm1}
\end{eqnarray}
with $\lambda=0,\pm 1$ for $\omega_{\lambda} = e^{i\frac{2\pi}{3}\lambda}$. The four remaining eigenvectors are obtained from the ones stated above by applying the spin flip $\ket{\uparrow} \leftrightarrow \ket{\downarrow}$ to all three spins. The fact that the eigenstates of $\hat{\chi}$ with nonzero eigenvalues are equivalent to the $W$ state~\cite{Cruz19} up to local operations has resulted in the proposal of the scalar spin chirality as a genuine tripartite entanglement witness~\cite{tsomokos08}. 



One of the motivations to explore quantum circuits to measure the scalar spin chirality  $\hat{\chi}$ is the prominent role it plays in various quantum many-body systems. In quantum magnetism, $\hat{\chi}$ has been proposed as a non-trivial order parameter that breaks parity ($P$) and time reversal ($T$) symmetries but preserves $PT$ symmetry in otherwise disordered chiral spin liquids~\cite{wen1989}. Moreover, in the strong-coupling limit where interactions prevent two electrons from occupying the same site, $\hat{\chi}$ is the orbital moment operator of circulating electrons in three-site loops, thus coupling linearly to an applied magnetic field \cite{sen95}.

In classical magnetism, where spins are described by classical vectors $\vec{m}_i=S\vec{n}_i$, with $||\vec{n}_i|| = 1$, 
the   scalar spin chirality is defined as
\begin{equation}
    \chi^{\rm cl}_{i,j,k} = S^3
    \vec{n}_i\cdot\left(\vec{n}_j\times\vec{n}_k\right),
\label{eq:clchir}
\end{equation}
providing a metric for the non-coplanarity of spin textures, such as  skyrmions~\cite{fert17} and frustrated ferromagnets~\cite{taguchi01}.
Recent works~\cite{roldan15,gauyacq19,ochoa19,sotnikov21,mazurenko23} have addressed a quantum description, beyond the broken-symmetry paradigm, of some of these non-coplanar magnetic configurations, identifying the quantum version of the scalar spin chirality (see Eq. (\ref{eq:chirality})) as the proper order parameter~\cite{sotnikov21}. 

The digital quantum simulation of chiral spin states encompasses two main lines of research: the synthesis of chiral logic gates to prepare and simulate the dynamics of chiral spin states~\cite{ma17,wang19}, and the development of measurement schemes to identify chiral spin states~\cite{Sotnikov22, mazurenko23}. Our work contributes towards the latter. While Sotnikov \textit{et al.}~\cite{Sotnikov22} proposed a hardware-efficient approach to identify phase transitions associated with changes in the scalar spin chirality \cite{sotnikov21} by measuring in two or more random single-qubit bases and coarse-graining the outcomes to probe inter-scale dissimilarities, here we follow a physically-motivated strategy whereby the expectation value of the scalar spin chirality is computed. 

As a first step in our analysis, we treat  classical and quantum spin chirality on equal footing by realizing that broken-symmetry magnetism can be described in terms of product states, 
\begin{eqnarray}
    \ket{\Psi}_{br.symm.}= |\vec{n}_1\rangle\otimes|\vec{n}_2\rangle \otimes ... \otimes |\vec{n}_N\rangle,
    \label{eq:BS}
\end{eqnarray}
where $|\vec{n}_i\rangle$ is a Bloch vector that describes, up to a global phase, any arbitrary spin-$\frac{1}{2}$ state. The unit vector $\vec{n}_i$ is the same as the one that appears in Eq. (\ref{eq:clchir}), so, for any product state of the form given by Eq. (\ref{eq:BS}), the classical scalar spin chirality (see Eq. \ref{eq:clchir}) coincides with the expectation value of its quantum version (see Eq. (\ref{eq:chirality})), ignoring the factor $\frac{4}{\sqrt{3}}$. The observable stated in Eq. (\ref{eq:chirality}) therefore allows for a unified description of chirality in both classical and quantum spin-$\frac{1}{2}$ systems. 

As an order parameter, the quantum scalar spin chirality $\hat{\chi}_{ijk}$ identifies features with no classical counterpart. Indeed, $\hat{\chi}_{ijk}$ takes nonzero values not only for separable states with non-collinear vectors $(\vec{n}_i, \vec{n}_j, \vec{n}_k)$ (see Eq. \ref{eq:BS}) 
but also for entangled states such as the eigenstates stated in Eq. (\ref{eq:pm1}) with $\lambda = \pm 1$. For the latter, it is not possible to separate the states of the three spins-$\frac{1}{2}$, but we can nonetheless compute the expectation values of $\hat{S}^{x}_{\alpha}$, $\hat{S}^{y}_{\alpha}$, and $\hat{S}^{z}_{\alpha}$ at sites $\alpha = i, j, k$ and construct three vectors of the form $(\langle \hat{S}^{x}_{\alpha} \rangle, \langle \hat{S}^{y}_{\alpha} \rangle, \langle \hat{S}^{z}_{\alpha} \rangle) \equiv \vec{s}_{\alpha}$. Unlike the classical case, $(\vec{s}_i, \vec{s}_j, \vec{s}_k)$ need not be non-collinear for $\hat{\chi}_{ijk} \neq 0$. For example, for $\ket{\psi} = \frac{1}{\sqrt{3}}(\ket{\uparrow_1 \downarrow_2 \downarrow_3} + e^{i 2\pi/3} \ket{\downarrow_1 \uparrow_2 \downarrow_3} + e^{-i 2\pi/3} \ket{\downarrow_1 \downarrow_2 \uparrow_3})$, the quantum scalar spin chirality is nonzero --- $\langle \psi | \hat{\chi}_{123} | \psi \rangle = 1$ --- but the three vectors are equal --- $\vec{s}_{1} = \vec{s}_{2} = \vec{s}_{3} = (0,0,-1/6)$.  


\begin{figure}[t]
\includegraphics[width=\linewidth]{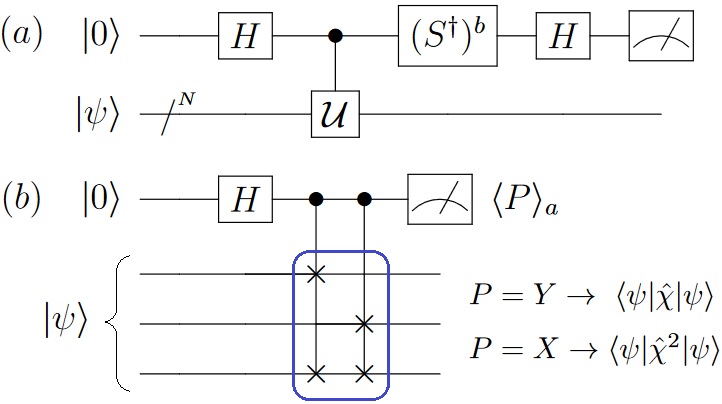}
\caption{Measurement of scalar spin chirality $\hat{\chi}$ via Hadamard test. (a) Scheme of Hadamard test. Setting $b = 0$ ($b = 1$) results in measurement of ancilla in X (Y) basis, the outcome of which is the real (imaginary) part of $\langle \psi | \mathcal{U} | \psi \rangle$. (b) Adaptation of Hadamard test by setting $\mathcal{U} = e^{-i \frac{2\pi}{3} \hat{\chi}}$, as highlighted by blue solid-line box. As detailed in Eq. (\ref{eq:H_test_outcomes}), measuring the ancilla in the Y (X) basis results in $\langle \psi | \hat{\chi} | \psi \rangle$ ($\langle \psi | \hat{\chi}^2 | \psi \rangle$). The scheme shows the three qubits of $\ket{\psi}$ that undergo a nontrivial action, but $\ket{\psi}$ can be of arbitrary size. $\ket{\psi}$ may be entangled or separable; the latter case corresponds to the cycle test~\cite{oszmaniec21}.}
\label{fig:H_test}
\end{figure}

We now describe a quantum routine based on the Hadamard test (see Fig. \ref{fig:H_test}) to compute the expectation value of the scalar spin chirality for a general state $\ket{\psi}$ of $N \geq 3$ spins-$\frac{1}{2}$, either separable or entangled. Using $\hat{\chi}^3 = \hat{\chi}$, the unitary $\mathcal{U}_{\tau} \equiv e^{-i \hat{\chi} \tau}$ ($\tau \in \mathbb{R}$) becomes
\begin{equation}
    \mathcal{U}_{\tau} \equiv e^{-i \hat{\chi} \tau}  = 1 + (\cos \tau -1) \, \hat{\chi}^2 - i \sin \tau \; \hat{\chi}.
\label{eq:comp_exp_chi}
\end{equation}
Setting $\tau = \frac{2\pi}{3}$ and plugging $\mathcal{U}_{\tau}$ into the Hadamard test (Fig. \ref{fig:H_test}(a)), the expectation values of $\hat{\chi}$ and $\hat{\chi}^2$ are obtained from measuring the ancillas in the Y and X bases:
\begin{equation}
    \begin{aligned}
    & \langle \psi | \hat{\chi} | \psi \rangle = - \frac{2}{\sqrt{3}} \, \langle Y \rangle_{a}, \\
    & \langle \psi | \hat{\chi}^2 | \psi \rangle = \frac{2}{3} (1 - \langle X \rangle_{a} ).
    \end{aligned}
\label{eq:H_test_outcomes}
\end{equation}

Importantly, this particular choice of $\tau = \frac{2\pi}{3}$ ensures the unitary operator $\mathcal{U}_{\tau}$ can be realized through the quantum circuit shown within the blue solid-line box in Fig. \ref{fig:H_test}(b). Ignoring qubit connectivity constraints, the three-qubit unitary $\mathcal{U}_{\tau = 2\pi/3}$ can be decomposed in terms of only $6$ \textsc{cnot} gates. The shallow character of this decomposition is even more pronounced once we control the two \textsc{swap} gates with the ancillary qubit in the Hadamard test: the total \textsc{cnot} count is $14$ \cite{Cruz23}, far below the maximum $100$ \textsc{cnot}s for a four-qubit operation~\cite{Krol22}.

Another interesting implication of this choice of $\tau = \frac{2\pi}{3}$ is the connection to the cycle test~\cite{oszmaniec21}, which has been recently proposed as a way to measure the unitary-invariant properties of a set of states, which characterize their relative geometrical orientation. In fact, the circuit in Fig. \ref{fig:H_test}(b) is precisely the cycle test in the case of a separable state $\ket{\psi} = \ket{\vec{n}_1} \otimes \ket{\vec{n}_2} \otimes \ket{\vec{n}_3}$. For such separable state inputs, the expectation value of the scalar spin chirality can be written as the imaginary part of the Bargmann invariant~\cite{Bargmann64}:
\begin{equation}
    \langle \psi | \hat{\chi} | \psi \rangle = \frac{2}{\sqrt{3}} {\rm Im} \left\{
    \langle \vec{n}_1|\vec{n}_2\rangle 
    \langle \vec{n}_2|\vec{n}_3\rangle 
    \langle \vec{n}_3|\vec{n}_1\rangle \right\}.
\end{equation}
This connection is not a coincidence: for the case of three single-qubit (i.e., spin-$\frac{1}{2}$) states, ${\rm Im} \left\{
    \langle \vec{n}_1|\vec{n}_2\rangle 
    \langle \vec{n}_2|\vec{n}_3\rangle 
    \langle \vec{n}_3|\vec{n}_1\rangle \right\} \neq 0$ if an only if $\vec{n}_1 \cdot (\vec{n}_2 \cross \vec{n}_3) \neq 0$, which is nothing more than the condition for a classical state to be chiral. 
Naturally, for entangled states, this geometric interpretation is no longer valid, but the cycle test can still probe $\hat{\chi}$.

We now test our method with a multi-spin-$\frac{1}{2}$ entangled state. We choose a spin-wave (i.e., one-magnon) state in a one-dimensional lattice~\footnote{The generalization to an arbitrary Bravais lattice is straightforward.},
\begin{equation}
    |q\rangle = \frac{1}{\sqrt{N}} \sum_{n=1}^{N} e^{i q n} \hat{S}^{-}_{n} \ket{\uparrow \uparrow ... \uparrow},
\end{equation}
where $N$ is the number of sites, $\hat{S}^{-}_{n}$ is the spin-lowering operator at site $n$, and $q=\frac{2\pi}{N} m$ is the wavenumber of the magnon, with $m \in \{ 0, 1, 2, ..., N-1 \}$.
After some straightforward algebra we find that
\begin{equation}
\langle q |\hat{\chi}_{n_1,n_2,n_3}|q\rangle = 
\frac{2 [ \sin(q \Delta_{21} ) +\sin(q\Delta_{32}) + \sin(q \Delta_{13} ) ] }{N\sqrt{3}},
\label{eq:analytical}
\end{equation}
where $\Delta_{ij} \equiv n_i - n_j$ and $n_1,n_2,n_3$ label arbitrary sites. For $N=3$ the spin-wave states with $q=0, \pm \frac{2\pi}{3}$ can be identified with the aforementioned eigenstates of the scalar spin chirality operator with eigenvalues $\lambda = 0, \pm 1$ (see Eq. (\ref{eq:pm1})), in which case we obtain $\langle q |\hat{\chi}_{1,2,3}|q\rangle= \frac{3}{2\pi} q $.

\begin{figure}[t]
\includegraphics[width=\linewidth]{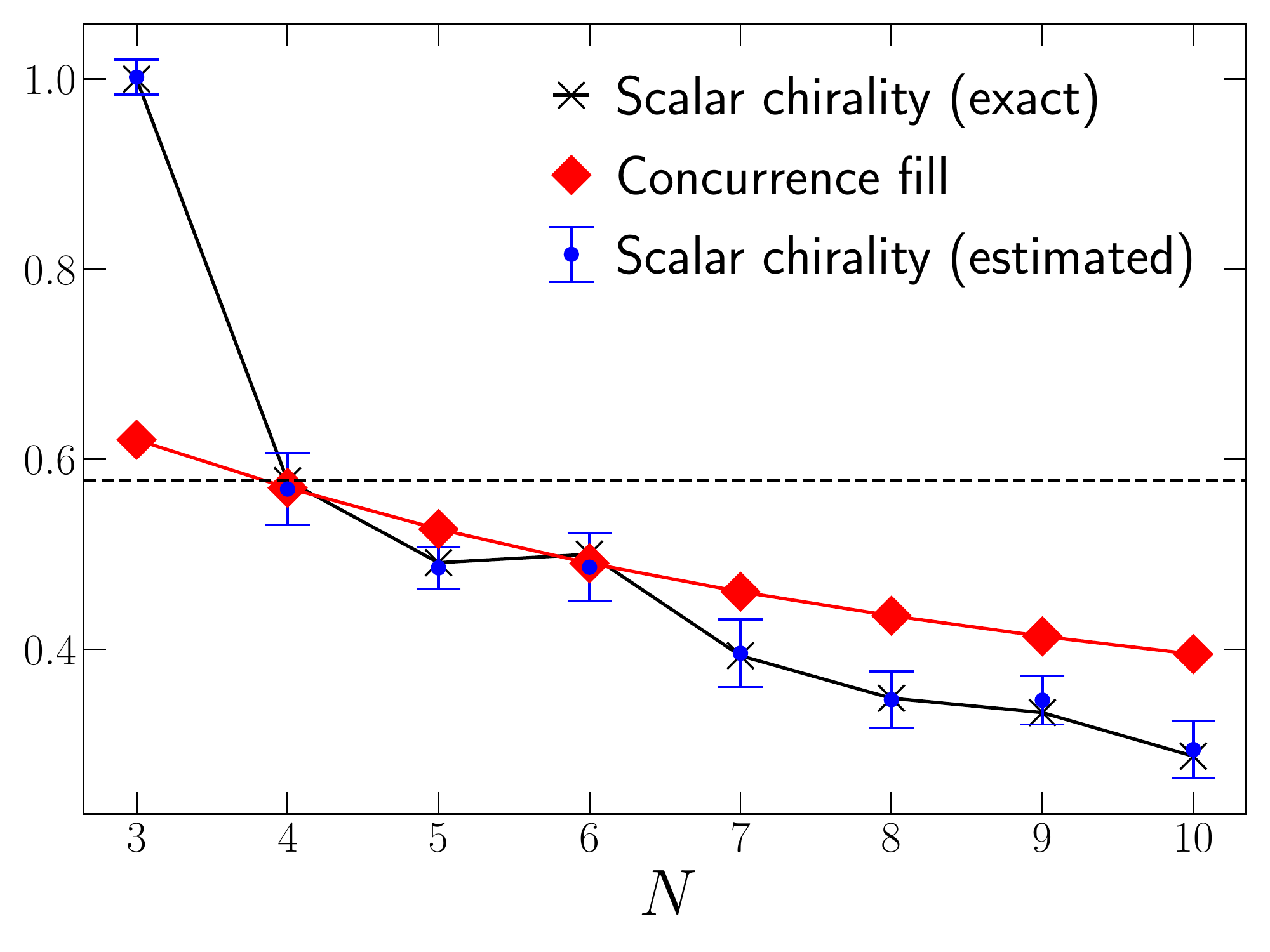}
\caption{Estimation of expectation value of scalar spin chirality $\hat{\chi}$ at trio of sites of spin-wave states defined on $N$-site ring. The maximal value of $\hat{\chi}$ across all trios of spins-$\frac{1}{2}$ and all $N$ spin-wave states for a given number of sites $N$ is shown in black, following Eq. (\ref{eq:analytical}). The same expectation values were estimated using the Hadamard test in Fig. \ref{fig:H_test} with 10,000 samples for each $N$ and the results are shown in blue. The horizontal dashed line marks the value above which $\hat{\chi}$ identifies genuine tripartite entanglement~\cite{tsomokos08}. Although the scalar spin chirality is only above this threshold for $N = 3$, the concurrence fill~\cite{Xie21}, which is a measure of tripartite entanglement, was also computed, confirming the existence of tripartite entanglement for all system sizes considered in the plot. The black and red solid lines were added to emphasize the decay of $\hat{\chi}$ and the concurrence fill with the system size $N$.}
\label{fig:spin_waves}
\end{figure}

The maximal values of the scalar spin chirality evaluated at any trio of spins-$\frac{1}{2}$ for spin-wave states defined on $N \in [3,10]$ sites are shown in black in Fig. \ref{fig:spin_waves}. The same expectation values were estimated via the Hadamard-test-based method depicted in Fig. \ref{fig:H_test} using a total of 10,000 samples for each $N$, as presented in blue in Fig. \ref{fig:spin_waves}. The spin-wave states were initialized via a generic state preparation method that exploits the Schmidt decomposition of the state~\cite{Plesch11}. The standard spin-$\frac{1}{2}$-to-qubit encoding $\{ \ket{\uparrow} \leftrightarrow \ket{0}, \ket{\downarrow} \leftrightarrow \ket{1} \}$ was assumed. The horizontal dashed line marks the value of $\frac{1}{\sqrt{3}}$~\footnote{According to Ref. \cite{tsomokos08}, the threshold above which the scalar spin chirality guarantees the existence of genuine tripartite entanglement is $2$. However, a different normalization of the scalar spin chirality was adopted --- one for which the nonzero eigenvalues are $\pm 2 \sqrt{3}$.}, above which the scalar spin chirality is a witness of genuine tripartite entanglement~\cite{tsomokos08}. Although the scalar spin chirality only guarantees the existence of genuine tripartite entanglement in spin-wave states for $N = 3$ sites, the nonzero value of the concurrence fill~\cite{Xie21} confirms the presence of tripartite entanglement for larger $N$, though it seems to vanish for a sufficiently large size. This contrasts with the macroscopic bipartite entanglement of magnons in the thermodynamic limit reported in the literature~\cite{Morimae05,Zou20}.

\begin{figure}[t]
\includegraphics[width=\linewidth]{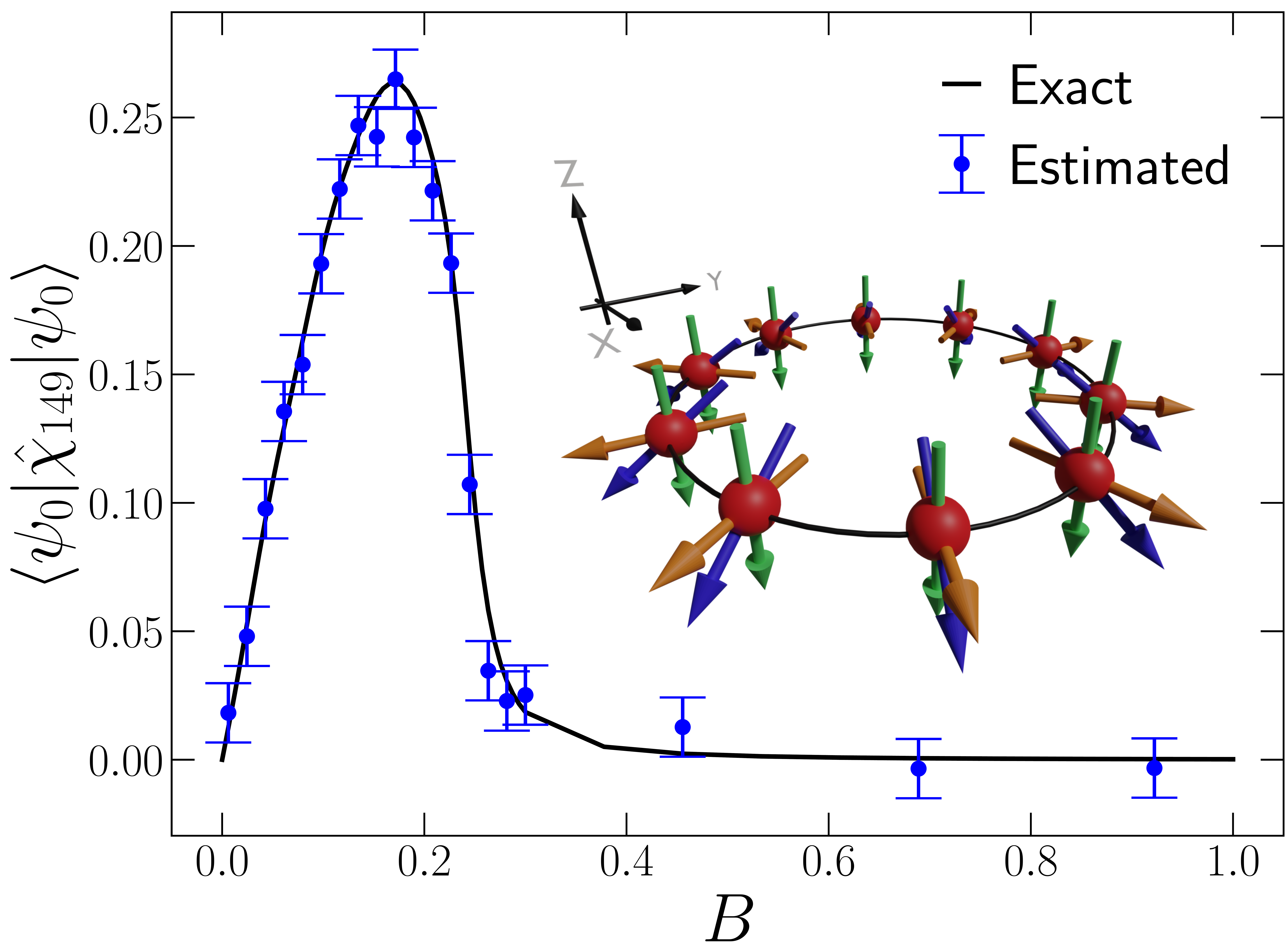}
\caption{Scalar spin chirality $\hat{\chi}$ for trio of
spins-$\frac{1}{2}$ at sites 1, 4, and 9 of ring with $N = 10$ sites for ground state $\ket{\psi_0}$ of Hamiltonian given in Eq. (\ref{eq:Hamiltonian}) with $J = 1$, $\vec{D} = J \tan(2\pi/N) \hat{z}$ and $B' = -0.1$. $\ket{\psi_0}$ was found via exact diagonalization and $\langle \psi_0 | \hat{\chi}_{149} | \psi_0 \rangle$ was computed numerically (black solid line). The same expectation value was computed via the Hadamard test (Fig. \ref{fig:H_test}) using 10,000 samples for each value of $B$. In the inset figure, we show the spin configurations, obtained by computing the expectation value of $\vec{S} = (\hat{S}^{x}_{i}, \hat{S}^{y}_{i}, \hat{S}^{z}_{i})$ at every site $i$ of the ring, for $B = 0$ (orange), $B = 0.17$ (blue), and $B=1$ (green). The first and third spin configurations are coplanar while the second is non-coplanar, as captured by $\hat{\chi}$.
}
\label{fig:chi_FM_DM_B}
\end{figure}

We have also estimated the scalar spin chirality of a trio of spins-$\frac{1}{2}$ in the ground state $\ket{\psi_0}$ of 
\begin{equation}
\hat{\mathcal{H}} = -J \sum_{n=1}^{N}\vec{S}_n\cdot\vec{S}_{n+1} + \vec{D} \cdot\sum_n
\vec{S}_n\times\vec{S}_{n+1} + B' \hat{S}_1^x + B \sum_{n = 1}^{N} \hat{S}_i^z, 
\label{eq:Hamiltonian}
\end{equation}
with $\vec{D} = D \vec{z}$ on a ring of $N=10$ sites. Fig. \ref{fig:chi_FM_DM_B} shows the results for a trio of spins at sites $1$, $4$ and $9$ of the ring with Hamiltonian parameters $J = 1$, $D = J \tan (2\pi/N)$ and $B' = -0.1$, with $B$ varied across the range $[0,1]$. For $B = 0$, the (ferromagnetic) Heisenberg and the Dzyaloshinskii–Moriya (DM) interactions give rise to a ground state that is a linear superposition of spin spiral states in the $xy$-plane with different phases. The chosen value for $D$ ensures that a period of the spin spirals is completed upon covering all $N$ sites. The nonzero value of $B'$ leads to a broken-symmetry state with a single spiral configuration in the ground state~\footnote{Classically, the competition between the ferromagnetic Heisenberg term (which favors the alignments of the spins) and the DM interaction (which promotes a perpendicular orientation between them) gives rise to a spin spiral configuration in the plane perpendicular to $\vec{D} = D\vec{z}$. Specifically, the classical spin vector at site $i$ is $(\cos (\theta i), \sin (\theta i), 0)$, with $\tan \theta = \frac{D}{J}$. However, a global rotation of all spins does not affect the energy of this spin spiral configuration, so any configuration with a local spin vector $(\cos (\theta i + \phi), \sin (\theta i + \phi), 0)$, with $\phi \in \mathbb{R}$, is also found in the ground state manifold. The additional Zeeman term generated by a local magnetic field, $B' S^{x}_{1}$, breaks this degeneracy explicitly, thus ensuring that the classical ground state at $B = 0$ is a single spin spiral.}. As $B$ is increased, a non-coplanar spin configuration arises, as attested by the nonzero value of the scalar spin chirality. For a sufficiently high $B$ the three spins-$\frac{1}{2}$ become aligned with the external magnetic field, and therefore the scalar spin chirality vanishes again.

We now compare this scheme to the default strategy of directly measuring the scalar spin chirality operator, as far as the total number of state preparation instances and single-qubit measurements are concerned. In order to directly measure the scalar spin chirality, it must be expanded in the Pauli basis:
\begin{equation}
    \hat{\chi} = \frac{XYZ + YZX + ZXY - XZY - YXZ - ZYX}{2 \sqrt{3}}.
    \label{eq:Pauli_basis}
\end{equation}
The standard approach amounts to measuring each Pauli string separately. Assuming that we want the error to be bounded by $\epsilon$, the total number of trials required is at most $N_{\textrm{Total}} = \frac{3}{\epsilon^2}$ \footnote{Using the expansion of $\hat{\chi}$ in the Pauli basis $\{ P_i \}$ (see Eq. (\ref{eq:Pauli_basis})) and the fact that the $N_{\textrm{Total}}$ trials are distributed evenly across the $6$ Pauli strings, the variance of $\langle \psi | \hat{\chi} | \psi \rangle \equiv \langle \hat{\chi} \rangle$ for a given state $\ket{\psi}$ is $\textrm{Var}(\langle \hat{\chi} \rangle) = \sum_{i = 1}^{6} \left( \frac{1}{2 \sqrt{3}} \right)^2 \frac{\textrm{Var}(\langle P_i \rangle)}{N_{\textrm{Total}}/6} \leq \frac{3}{N_{\textrm{Total}}}$, where in the last step we used the fact that $\textrm{Var}(\langle P_i \rangle) \leq 1$ for any state $\ket{\psi}$ and any Pauli string $P_i$.}. For the Hadamard test, in turn, achieving a precision of $\epsilon$ requires at most $N_{\textrm{Total}} = \frac{4}{3 \epsilon^2}$. Hence, for the same precision, the Hadamard test results in $\frac{9}{4} = 2.25$ fewer state preparation instances than the standard approach. If we consider instead the number of single-qubit measurements, then the default option involves $3$ single-qubit measurements for each Pauli string as opposed to the sole measurement of the ancilla in the Y basis in each trial of the Hadamard test. Hence, the Hadamard test yields $\frac{27}{4} = 6.75$ fewer single-qubit measurements than the default method. When implemented in $M$ trios of spins-$\frac{1}{2}$ of a multi-spin-$\frac{1}{2}$ state, the total number of measurements is reduced by factor of $6.75 \, M$. This is particularly relevant in near-term quantum hardware, where the measurement operations have the highest error rates~\cite{Funcke22} and execution times~\cite{Cheng23}. 
%
%
%

Finally, we consider the use of quantum phase estimation with a single ancillary qutrit~\cite{tonchev16} to measure, in one shot, the eigenvalue of an input eigenstate of the scalar spin chirality. Since only one ancilla is required, the corresponding quantum circuit is essentially the one shown in Fig. \ref{fig:H_test}(a) with $\mathcal{U} = e^{-i \frac{2\pi}{3} \hat{\chi}}$ and $b = 0$, but with two differences due to the use of a qutrit instead of a qubit as the ancilla. First, the Hadamard gate before the controlled-unitary (and its inverse after it) has to be replaced by a $3 \times 3$ quantum Fourier transform,
\begin{equation}
    \textrm{QFT}_3 = \frac{1}{\sqrt{3}} \begin{pmatrix}
     1 & 1 & 1 \\
     1 & w & w^2 \\
     1 & w^2 & w
    \end{pmatrix},
\end{equation}
where $w = e^{i \frac{2\pi}{3}}$. Second, the application of the unitary operation $\mathcal{U} = e^{-i \frac{2\pi}{3} \hat{\chi}}$ to the state $\ket{\psi}$ in the main register subject to the control of the ancillary qutrit in the computational basis $\ket{i}$ follows the relation
\begin{equation}
    \mathcal{O}_{\mathcal{U}} (\ket{i} \otimes \ket{\psi}) = \ket{i} \otimes \; \mathcal{U}^{i} \ket{\psi} ,
\end{equation}
with $i = 0, 1, 2$. 
For the three distinct eigenvalues of $\hat{\chi}$, having one of the corresponding eigenstates as the input state $\ket{\psi}$ leaves the ancillary qutrit in a different computational basis state with certainty:
\begin{equation*}
    \begin{aligned}
    & \hat{\chi} \ket{\psi} = 0 \quad \; \; \; \Leftrightarrow \mathcal{U} \ket{\psi} = \ket{\psi} \quad \; : \textrm{Ancilla measured in} \ket{0}, \\
    & \hat{\chi} \ket{\psi} = - \ket{\psi} \Leftrightarrow \mathcal{U} \ket{\psi} = w \ket{\psi} \; \, : \textrm{Ancilla measured in} \ket{1}, \\
    & \hat{\chi} \ket{\psi} = + \ket{\psi} \Leftrightarrow \mathcal{U} \ket{\psi} = w^2 \ket{\psi} : \textrm{Ancilla measured in} \ket{2}.
    \end{aligned}
\end{equation*}
The expectation value of the scalar spin chirality can be estimated as $\langle \psi | \hat{\chi} | \psi \rangle = P_2 - P_1$, where $P_i$ is the probability of measuring the ancillary qutrit in state $\ket{i}$.  

Importantly, if the main register is prepared in a linear superposition of states with different scalar spin chirality, once the QPE algorithm is executed, the wave function collapses onto the components indicated by the ancilla readout. This approach~\cite{lacroix20} can be used as a strategy to prepare states with well-defined scalar spin chirality in chosen triads of spins, in the same vein of the recently proposed algorithms to prepare the valence-bond-solid states~\cite{murta23} and the Gutzwiller wave function~\cite{murta21}. 

In summary, we have proposed quantum circuits to probe the scalar spin chirality for trios of spins-$\frac{1}{2}$ in a wave function defined on an arbitrary lattice. The first circuit is based on the Hadamard test with ${\cal U}=e^{-i \tau \hat{\chi}}$ as the reference unitary operator (see Fig. \ref{fig:H_test}). It maps the expectation value of either $\hat{\chi}$ or $\hat{\chi}^2$ onto the average value of $Y$ or $X$ for the ancillary qubit. For $\tau=\frac{2\pi}{3}$ the algorithm is identical to the cycle test~\cite{oszmaniec21}, but it works for both separable and entangled states. We have illustrated the application of this method to one-magnon states, as well as to the ground state of a toy model of a chiral ferromagnet. Finally, when the Hadamard test is implemented with a qutrit as the ancilla, it is identical to quantum phase estimation and enables the single-shot readout of the eigenvalues of $\hat{\chi}$ when the main register is prepared in a corresponding eigenstate. 

Our results pave the way towards the efficient digital quantum simulation of magnetic materials with chiral properties. In addition, our quantum schemes provide a connection between the scalar spin chirality formula that is widely used in classical descriptions of magnetism, which can be formulated in terms of product states, and its quantum upgrade, which also applies to entangled states. Future work will address the extension of our results to the determination of the scalar spin chirality of spin systems with $S>\frac{1}{2}$ and to the generalization of this concept to sets of more than three spins.




%
{\em Acknowledgements.} L. I. R. thanks the New Talents program of Funda\c{c}\~{a}o Calouste Gulbenkian (Portugal) for financial support. B.M. acknowledges financial support from Funda\c{c}\~{a}o para a Ci\^{e}ncia e a Tecnologia (FCT) --- Portugal through the PhD scholarship No. SFRH/BD/08444/2020. E.F.G. acknowledges support from FCT via project CEECINST/00062/2018, and from the Digital Horizon Europe project FoQaCiA, GA no. 101070558. J.F.R.  acknowledges financial support from 
 FCT (Grant No. PTDC/FIS-MAC/2045/2021),
 SNF Sinergia (Grant Pimag),
Generalitat Valenciana funding Prometeo2021/017
and MFA/2022/045,
and
 funding from
MICIIN-Spain (Grant No. PID2019-109539GB-C41).



\bibliographystyle{apsrev4-1}
\bibliography{biblio}{}

\end{document}


\title{Efficient Quantum Algorithm to probe spin chirality in a quantum computer}

\author{L. I. Reascos $^{1}$, B. Murta $^{2}$, E. Galvao $^2$,J. Fern\'{a}ndez-Rossier$^2$\footnote{On permanent leave from Departamento de F\'{i}sica Aplicada, Universidad de Alicante, 03690 San Vicente del Raspeig, Spain}$^,$\footnote{joaquin.fernandez-rossier@inl.int} }

\affiliation{$^1$Centro de F\'{i}sica das Universidades do Minho e do Porto, Universidade do Minho, Campus de Gualtar, 4710-057 Braga, Portugal }

\affiliation{
$^2$International Iberian Nanotechnology Laboratory (INL), Av. Mestre Jos\'{e} Veiga, 4715-330 Braga, Portugal }

\date{\today}

\begin{appendix}
\maketitle

\section{title}

\end{appendix}

\bibliography{biblio}